\documentclass[12pt]{article}
\pdfoutput=1

\font\blackboard=msbm10 at 12pt
\font\blackboards=msbm7
\font\blackboardss=msbm5
\newfam\black
\textfont\black=\blackboard
\scriptfont\black=\blackboards
\scriptscriptfont\black=\blackboardss

\begin{document}
\pagestyle{plain}
\setcounter{page}{1}

\baselineskip16pt

\begin{titlepage}

\begin{flushright}

\end{flushright}
\vspace{8 mm}

\begin{center}

{\Large \bf U-duality transformation of membrane on $T^{n}$ revisited \\}

\vspace{1cm}

{\large Shan Hu$^{a}$,
Tianjun Li$^{b,c,d}$}
\vspace{.5cm}

{\baselineskip 20pt \it

$^a$Department of Physics and Electronic Technology, Hubei University, Wuhan 430062, P. R. China \\

\vspace{2mm}
$^b$Key Laboratory of Theoretical Physics and
Kavli Institute for Theoretical Physics China (KITPC),
Institute of Theoretical Physics, Chinese Academy of Sciences,
Beijing 100190, P. R. China \\

\vspace{2mm}
$^c$School of Physical Sciences, University of Chinese Academy of Sciences,
Beijing 100049, P. R. China \\

\vspace{2mm}
$^d$School of Physical Electronics, University of Electronic Science and Technology of China,
Chengdu 610054, P. R. China \\

}

\end{center}

\vspace{.5cm}

\begin{abstract}

  The problem with the U-duality transformation of membrane on $T^{n}$ is recently addressed in [arXiv:1509.02915 [hep-th]]. We will consider the U-duality transformation rule of membrane on $T^{n}\times R$. It turns out that winding modes on $T^{n}$ should be taken into account, since the duality transformation may bring the membrane configuration without winding modes into the one with winding modes. With the winding modes added, the membrane worldvolume theory in lightcone gauge is equivalent to the $n+1$ dimensional super-Yang-Mills (SYM) theory in $\tilde{T}^{n}$, which has $SL(2,Z)\times SL(3,Z)$ and $SL(5,Z)$ symmetries for $n=3$ and $n=4$, respectively. The $SL(2,Z)\times SL(3,Z)$ transformation can be realized classically, making the on-shell field configurations transformed into each other. However, the $SL(5,Z)$ symmetry may only be realized at the quantum level, since the classical $5d$ SYM field configurations cannot form the representation of $SL(5,Z)$. 

\end{abstract}

\vspace{1cm}
\begin{flushleft}

\end{flushleft}
\end{titlepage}
\newpage

\section{Introduction}

It is well-known that string theory compactified on $T^{n}$ has the $O(n,n;Z)$ T-duality symmetry, which is manifested in the classical equations of motion for both background fields and the $2d$ sigma model. In \cite{d1}, it has been shown that string T-duality can be realized on the string worldsheet as a rotation of field equations into Bianchi identities. M theory compactified on $T^{n}$ also has the U-duality symmetry, which, when $n=3$ and $n=4$, are $SL(2,Z)\times SL(3,Z)$ and $SL(5,Z)$ respectively. Equations of motion for the $11d$ supergravity are U-duality invariant. It remains to find the U-duality transformation rule of membrane on the given supergravity background. A natural generalization of \cite{d1} in the membrane case is to realize the U-duality transformation as the rotation of field equations into Bianchi identities on the $3$ dimensional worldvolume \cite{dd1}. However, the approach has some difficulties as discussed recently in \cite{ddd1}. Concretely, when $n=3$, the $SL(2,R)\times SL(3,R)$ transformation is well-defined, but for $n=4$, only a subgroup $GL(4,R) \times  R^{4}$ can be realized.

Nevertheless, because M theory compactified on $T^{n}$ is U-duality symmetric, there must be a U-duality transformation rule for membrane. In this note, we will reconsider this problem. Different from the approaches in \cite{dd1,ddd1}, where the worldvolume theory of membrane is covariant, we will impose the lightcone gauge, and then, after the discrete regularization, membrane worldvolume theory becomes $U(\infty)$ Matrix model. Another difference is that in \cite{dd1,ddd1}, membrane worldvolume, including the time direction, is totally embedded in $T^{n}$, but in our discussion, membrane lives in $T^{n} \times R$ with the time direction identified as $R$. As a result, some identities in \cite{ddd1} are not valid here. For example, membrane in $T^{3}$ is a topological object with the winding current proportional to the momentum current due to some algebraic relations \cite{ddd1}. On the other hand, membrane in $T^{3}\times R$ carries dynamical degrees of freedom, while the winding current and the momentum current are independent.

Branes in compact space contain more degrees of freedom than branes in non-compact space, coming from the winding modes on non-trivial 1-cycles. For example, the worldvolume theory of a single $D2$ in $10d$ non-compact spacetime is the $3$ dimensional $U(1)$ super-Yang-Mills (SYM) theory. If $n$ transverse dimensions are compactified to $T^{n}$, open string winding modes along $T^{n}$ should also be added so that the theory is equivalent to the $(3+n)$ dimensional $U(1)$ SYM theory. Duality transformation usually makes branes without winding modes transformed into branes with winding modes. As an instance, let us consider the type IIA theory compactified on $T^{4}\sim x_{1}\times x_{2}\times x_{3}\times x_{4}$ and a $D2$ wrapping $x_{1}\times x_{2}$ with no winding modes. If $D2$ is not translation invariant along $x_{1}$ and $x_{2}$, after two successive T-duality transformations along $x_{1}$ and $x_{2}$, it will become a $D0$ with winding modes along $x_{1}\times x_{2}$ included. The further T-duality transformations along $x_{3}$ and $x_{4}$ give $D2$, which wraps $x_{3}\times x_{4}$ and is translation invariant along $x_{3}$ and $x_{4}$, still including the winding modes along $x_{1}\times x_{2}$. Similarly, we may expect that for membrane in $T^{n}$, winding modes should also be added and the U-duality transformation may make a configuration without winding modes transform into the one with winding modes.

Membrane in $T^{n}\times R$ in lightcone gauge with all winding modes included is equivalent to the $(n+1)$-dimensional $U(\infty)$ SYM theory in the dual $\tilde{T}^{n}\times R$. Without the winding modes, membrane configuration is mapped to the 0-mode of the SYM field. We will consider the $SL(2,Z)\times SL(3,Z)$ and $SL(5,Z)$ transformations for the $4d$ and $5d$ SYM theories, respectively. It turns out that the duality transformation does not always convert the $0$-mode into the $0$-mode, so starting from the membrane configuration without winding modes, the U-dual configuration may involve winding modes. For the $SL(2,Z)$ transformation of the $4d$ YM fields, based on the loop space formulation, \cite{1g, 7hg} gives a prescription converting the classical on-shell fields into each other. On the other hand, the $SL(5,Z)$ transformation of the $5d$ SYM fields cannot always be realized classically. The $5d$ SYM theory in $T^{4}\times R$, with all instanton configurations taken into account, could be taken as the $6d$ $(2,0)$ theory in $T^{5}\times R$, which is supposed to be $SL(5,Z)$ symmetric. Each $5d$ SYM field configuration carries the definite instanton number, or in other words, the definite $P^{5}$ momentum, and thus could not form the representation of $SL(5,Z)$, except for the $SL(4,Z)$ subgroup on $T^{4}$. Nevertheless, the $SL(5,Z)$ symmetry may be realized at the quantum level, for example, the partition function of the $5d$ SYM theory on $T^{4}\times R$ may be $SL(5,Z)$ invariant, although the explicit verification is still lacking.

The rest of the paper is organized as follows. In Section 2, we review the $SO(n,n)$ transformation for strings on $T^{n}$, based on \cite{d1}. In Section 3, we comment on the problems for the U-duality transformation of membrane on $T^{n}$, following \cite{ddd1}. In section 4, we review the matrix theory description of the membrane on $T^{n} \times R$. In Section 5, we discuss the U-duality transformation of membrane on $T^{3} \times R$. In Section 6, we consider the U-duality transformation of membrane on $T^{4} \times R$. The conclusion is in Section 7.

\section{$SO(n,n)$ transformation for string on $T^{n}$}

In \cite{d1}, it was pointed out that string T-duality originates from transforming field equations into Bianchi identities on the string worldsheet. Consider the worldsheet theory of the string on $T^{n}$ with the constant background fields $g_{\mu\nu}$ and $b_{\mu\nu}$, the equations of motion and the Bianchi identity can be interpreted as the conservation equations for the currents $p^{i}_{\mu}$ and $j^{i\mu}$: 
\begin{eqnarray}\label{1}
\nonumber  && 	\partial_{i}(g_{\mu\nu}\sqrt{-\gamma}\gamma^{ij}\partial_{j}x^{\nu}+b_{\mu\nu}\epsilon^{ij}\partial_{j}x^{\nu})=\partial_{i}p^{i}_{\mu}=0\;,\\
 && \partial_{i}(\epsilon^{ij}\partial_{j}x^{\mu})=\partial_{i}j^{i\mu}=0\;,
\end{eqnarray}
where $	\gamma_{ij}=\partial_{i}x^{\mu}\partial_{j}x^{\nu}g_{\mu\nu}$, $i=0,1$, $\mu=1,2,\cdots,n$, and
\begin{eqnarray}
  && 		p^{i}_{\mu}=g_{\mu\nu}\sqrt{-\gamma}\gamma^{ij}\partial_{j}x^{\nu}+b_{\mu\nu}\epsilon^{ij}\partial_{j}x^{\nu}\;,\\
 && 	j^{i\mu}=\epsilon^{ij}\partial_{j}x^{\mu}\;.
\end{eqnarray}
Integrations of $p^{i}_{\mu}$ and $j^{i\mu}$ give $P_{\mu}$ and $J^{\mu}$, the momentum and winding number of the string in $\mu$ direction. $P_{\mu}$ and $J^{\mu}$ are integers. Under the $SO(n,n;Z)$ T-duality transformation $\Lambda$, $(P,J)$ transforms as 
\begin{equation}     
\left(                 
\begin{array}{c}   
P \\  
J \\  
\end{array}
\right) \rightarrow \left(                 
\begin{array}{c}   
P' \\  
J' \\  
\end{array}
\right) = \Lambda          \left(                 
\begin{array}{c}   
P \\  
J \\  
\end{array}
\right) .                
\end{equation}
Although the exact symmetry is $SO(n,n;Z)$, one can nevertheless first consider the continuous group $SO(n,n;R)$, under which the infinitesimal transformation of $(P,J)$ is  
\begin{eqnarray}
\nonumber && 	\delta P_{\mu}= -A^{\alpha}_{\;\;\mu}P_{\alpha} + C_{\mu\beta} J^{\beta}\;
,\\
 && \delta J^{\nu}=B^{\nu\alpha} P_{\alpha} + A^{\nu}_{\;\;\beta} J^{\beta}\;.
\end{eqnarray}
A stronger requirement is that the current $(p^{i}_{\mu},j^{i\nu})$ transforms in the same way as the charge $(P_{\mu},J^{\nu})$:
\begin{eqnarray}\label{3s}
\nonumber && 	\delta p^{i}_{\mu}= -A^{\alpha}_{\;\;\mu}p^{i}_{\alpha} + C_{\mu\beta} j^{i\beta}\;
,\\
 && \delta j^{i\nu}=B^{\nu\alpha} p^{i}_{\alpha} + A^{\nu}_{\;\;\beta} j^{i\beta}\;.
\end{eqnarray}

The background fields $g_{\mu\nu}$ and $b_{\mu\nu}$ can be compactly written as a $2n \times 2n$ metric $G_{MN}$ 
\begin{equation}    
G_{MN}=\left(                
 \begin{array}{cc}   
g_{\mu\nu}+b_{\mu\alpha}g^{\alpha\beta}b_{\nu\beta} & b_{\mu\alpha}g^{\alpha\beta} \\ 
  g^{\alpha\beta}b_{\nu\beta}   & g^{\alpha\beta}\\  
 \end{array}
\right) \; .               
\end{equation}
The $SO(n,n)$ transformation rule for $G_{MN}$ is $G\rightarrow G'= \Lambda G \Lambda^{T}$, which gives 
\begin{eqnarray}\label{2s}
\nonumber && 	\delta g_{\mu\nu}=-A^{\rho}_{\;\;\mu}g_{\rho\nu}-A^{\rho}_{\;\;\nu}g_{\rho\mu}-b_{\mu\alpha}B^{\alpha\beta}g_{\beta\nu}-g_{\mu\alpha}B^{\alpha\beta}b_{\beta\nu}\;,\\
 && 	\delta b_{\mu\nu}=-A^{\rho}_{\;\;\mu}b_{\rho\nu}-A^{\rho}_{\;\;\nu}b_{\rho\mu}-b_{\mu\alpha}B^{\alpha\beta}b_{\beta\nu}-g_{\mu\alpha}B^{\alpha\beta}g_{\beta\nu}+C_{\mu\nu}\;.
\end{eqnarray}

From (\ref{3s}) and (\ref{2s}), $\delta \partial_{i}x^{\mu}$ is determined to be 
\begin{equation}\label{4ew}
	\delta \partial_{i}x^{\mu}=B^{\nu\alpha} (-g_{\alpha\nu}\sqrt{-\gamma}\epsilon_{ik}\gamma^{kj}\partial_{j}x^{\nu}+b_{\alpha\nu}\partial_{i}x^{\nu}) + A^{\nu}_{\;\;\beta} \partial_{i}x^{\beta}\;.
\end{equation}
With the equations of motion imposed, $\epsilon^{ij}\partial_{i}	\delta \partial_{j}x^{\mu}=0$, so (\ref{4ew}) can be integrated to give some $\delta x^{\mu}$. $\delta x^{\mu}$ together with (\ref{2s}) composes the $SO(n,n;R)$ transformation rule with the momentum and winding number current $(p^{i}_{\mu},j^{i\nu})$ transforming as (\ref{3s}).

\section{Problem for the U-duality transformation of membrane on $T^{n}$}

Similar to the string situation, it is naturally expected that the M theory U-dualities originate from transforming field equations into Bianchi identities on the membrane worldvolume \cite{dd1}. However, unless the target space has dimension D = p + 1, there is a problem that is identified as the non-integrability of the U-duality transformation assigned to the pull-back map \cite{ddd1}.

For the worldvolume theory of membrane on $T^{n}$ with the constant background fields $g_{\mu\nu}$ and $b_{\mu\nu\rho}$, the equations of motion and the Bianchi identity are
\begin{eqnarray}\label{101}
\nonumber  && 	\partial_{i}(g_{\mu\nu}\sqrt{-\gamma}\gamma^{ij}\partial_{j}x^{\nu}+\frac{1}{2}b_{\mu\nu\rho}\epsilon^{ijk}\partial_{j}x^{\nu}\partial_{k}x^{\rho})=\partial_{i}p^{i}_{\mu}=0\;,\\
 && 	\partial_{i}(\epsilon^{ijk}\partial_{j}x^{\mu}\partial_{k}x^{\nu})=\partial_{i}j^{i\mu\nu}=0\;.
\end{eqnarray}
(\ref{101}) can be interpreted as the conservation laws of the current $p^{i}_{\mu}$ and $j^{i\mu\nu}$, and the integration of which gives the momentum $P_{\mu}$ and membrane wrapping number $J^{\mu\nu}$, where $i=0,1,2$, $\mu=1,2,\cdots,n$.
$P_{\mu}$ and $J^{\mu\nu}$ are also integers.
\begin{eqnarray}
  && 		p^{i}_{\mu}=g_{\mu\nu}\sqrt{-\gamma}\gamma^{ij}\partial_{j}x^{\nu}+\frac{1}{2}b_{\mu\nu\rho}\epsilon^{ijk}\partial_{j}x^{\nu}\partial_{k}x^{\rho}\;,\\
 && 	j^{i\mu\nu}=\epsilon^{ijk}\partial_{j}x^{\mu}\partial_{k}x^{\nu}\;.
\end{eqnarray}

For $n=3$, there is the identity
\begin{equation}
g_{\mu\nu}\sqrt{-\gamma}\gamma^{ij}\partial_{j}x^{\nu}=-\frac{1}{2}\sqrt{-g}\epsilon_{\mu\nu\rho}\epsilon^{ijk}\partial_{j}x^{\nu}\partial_{k}x^{\rho}\;.
\end{equation}
Let 
\begin{equation}
	b_{\mu\nu\rho}=\sqrt{-g}\epsilon_{\mu\nu\rho}b\;, \;\;\;\;\;\;\;\;j^{i}_{\mu}=\frac{1}{2}\epsilon_{\mu\nu\rho}j^{i\nu\rho}\;,
\end{equation}
also note that 
\begin{equation}
	\epsilon^{ijk}\epsilon_{\mu\nu\rho}\partial_{j} x^{\nu} \partial_{k} x^{\rho}=-2(\det \partial x)\gamma^{ij} g_{\mu\nu}\partial_{j}x^{\nu}=-2|\gamma|^{1/2}|g|^{-1/2}\gamma^{ij} g_{\mu\nu}\partial_{j}x^{\nu}\;,
\end{equation}
$p^{i}_{\mu}$ and $j^{i}_{\mu}$ can be simplified as 
\begin{equation}
	p^{i}_{\mu}=-\sqrt{-\gamma}\gamma^{ij}g_{\mu\nu}(1+b)\partial_{j}x^{\nu}\;,\;\;\;\;\;\;\;\;\;j^{i}_{\mu}=\sqrt{-\gamma}\gamma^{ij}g_{\mu\nu}|g|^{-1/2}\partial_{j}x^{\nu}
\end{equation}
with 
\begin{equation}\label{ll}
	p^{i}_{\mu}=-(1+b)|g|^{1/2}j^{i}_{\mu}\equiv C j^{i}_{\mu}\;.
\end{equation}
The $GL(2,R)=SL(2,R)\times R$ transformation of $(	p^{i}_{\mu},j^{i}_{\mu})$ is required to be 
\begin{equation}
	\delta  \left(                
 \begin{array}{c}   
p^{i}_{\mu} \\ 
 j^{i}_{\mu}\\  
 \end{array}
\right)=\left(                
 \begin{array}{cc}   
\alpha & \beta \\ 
 \gamma  & -\alpha\\  
 \end{array}
\right)\left(                
 \begin{array}{c}   
p^{i}_{\mu} \\ 
 j^{i}_{\mu}\\  
 \end{array}
\right)+\lambda \left(                
 \begin{array}{c}   
p^{i}_{\mu} \\ 
 j^{i}_{\mu}\\  
 \end{array}
\right)\;,
\end{equation}
which can be satisfied when
\begin{equation}
	\delta C=\beta+2\alpha C-\gamma C^{2}\;,
\end{equation}
\begin{equation}\label{ews2}
	\delta (\partial_{i}X^{\mu})=(C\gamma-\alpha+\lambda)\partial_{i}X^{\mu}\;.
\end{equation}
(\ref{ews2}) is of course integrable and then the defined transformation is consistent.

For $n=4$, the U-duality symmetry of M theory is $SL(5,Z)$. Let us define $K_{\mu5}\equiv P_{\mu}$, $K_{\mu\nu}\equiv \frac{1}{2}\epsilon_{\mu\nu\rho\sigma}J^{\rho\sigma}$, $K\equiv(K_{\mu5},K_{\mu\nu})$ forms a $5 \times 5$ antisymmetric matrix, which, under the action of $\Lambda \in SL(5,Z)$, transforms as 
\begin{equation}\label{cf1}
	K \rightarrow K'= \Lambda K \Lambda^{T}\;.
\end{equation}
If one requires the worldvolume Lorentz invariance, the current $(p^{i}_{\mu},j^{i\nu\rho})$ will transform in the same way as the charge $(P_{\mu},J^{\nu\rho})$, and if the $SL(5,Z)$ symmetry is relaxed to $SL(5,R)$, the infinitesimal transformation of $(p^{i}_{\mu},j^{i\nu\rho})$ will be 
\begin{eqnarray}\label{1qc}
\nonumber && 	\delta p^{i}_{\mu}= (-A^{\alpha}_{\;\;\mu}+\frac{3}{4}A \delta^{\alpha}_{\mu})p^{i}_{\alpha} + \frac{1}{2}B_{\mu\beta\gamma} j^{i\beta\gamma}\;
,\\
 && \delta j^{i\nu\rho}=C^{\nu\rho\alpha} p^{i}_{\alpha} + (2A^{[\nu}_{\;\;[\beta} \delta^{\rho]}_{\gamma]}-\frac{1}{2}A\delta^{\nu\rho}_{\beta\gamma})j^{i\beta\gamma}\;.
\end{eqnarray}
$g_{\mu\nu}$ and $b_{\mu\nu\rho}$ can also be assembled into a symmetric $5 \times 5$ matrix $G_{MN}$ with $M,N=1,\cdots,5$,
\begin{eqnarray}
\nonumber  && G_{\mu\nu}= g^{-2/5}g_{\mu\nu}\;, \;\;\;\;\; G_{\mu5}=G_{5\mu}= \frac{1}{3!} g^{-2/5} g_{\mu\alpha}\epsilon^{\alpha\beta\gamma\delta}b_{\beta\gamma\delta}\;,\\
 && 	G_{55}=g^{3/5}(1+\frac{1}{3!}b^{2})\;,\;\;\;\;\;b^{2}\equiv b_{\mu\nu\rho}b^{\mu\nu\rho}\;.
\end{eqnarray}
Under the $SL(5,R)$ transformation, $G\rightarrow G'=\Lambda G \Lambda^{T}$,
\begin{eqnarray}\label{1qq}
\nonumber && 	\delta g_{\mu\nu}= -2A^{\sigma}_{\;\;(\mu}g_{\nu)\sigma}+\frac{5}{6}(A+\frac{2}{15}C^{\alpha\beta\gamma}b_{\alpha\beta\gamma})g_{\mu\nu}-C^{\alpha\beta\gamma}g_{\gamma(\mu}b_{\nu)\alpha\beta}\;
,\\
 && \delta b_{\mu\nu\rho}=-3A^{\sigma}_{\;\;[\mu}b_{\nu\rho]\sigma}+\frac{5}{4} (A-\frac{2}{15}C^{\alpha\beta\gamma}b_{\alpha\beta\gamma})b_{\mu\nu\rho} +B_{\mu\nu\rho} +C^{\alpha\beta\gamma}g_{\mu\alpha}g_{\nu\beta}g_{\rho\gamma}\;,
\end{eqnarray}
where
\begin{equation}
\epsilon=\frac{1}{\sqrt{-\gamma}}\epsilon^{lmn}\partial_{l}x^{\mu}\partial_{m}x^{\nu}\partial_{n}x^{\rho}g_{\mu\alpha}g_{\nu\beta}g_{\rho\gamma}B^{\alpha\beta\gamma}\;.
\end{equation}
From (\ref{1qc}), (\ref{1qq}) and the constraint $\gamma_{ij}=\partial_{i}x^{\mu}\partial_{j}x^{\nu}g_{\mu\nu}$, $\delta \partial_{i}x^{\mu}$ is determined to be 
\begin{equation}
	\delta \partial_{i}x^{\mu}=[A^{\mu}_{\;\;\sigma}-\frac{1}{4}(A+\frac{1}{3}C^{\alpha\beta\gamma}b_{\alpha\beta\gamma}+\frac{1}{3}\epsilon)\delta^{\mu}_{\sigma}+\frac{1}{2}C^{\mu\alpha\beta}b_{\alpha\beta\sigma}]\partial_{i}x^{\sigma}+\frac{1}{2}C^{\mu}_{\;\;\nu\rho}\frac{\gamma_{ij}j^{j\nu\rho}}{\sqrt{-\gamma}}\;.
\end{equation}
$\epsilon^{ijk}\partial_{j}	\delta \partial_{k}x^{\mu}=0$ does not necessarily hold when $C^{\alpha\beta\gamma}\neq 0$, even if the equations of motion is imposed. This just indicates that we cannot find a $\delta x^{\mu}$ with $(p^{i}_{\mu},j^{i\nu\rho})$ transforming as (\ref{1qc}).

A possible reason is that maybe (\ref{1qc}) is a too strong requirement. We may only need the charge $K_{MN}$ transforming as a $5 \times 5$ antisymmetric matrix like that in (\ref{cf1}). In fact, we do have examples for which, the U-duality transformation is implemented respecting (\ref{cf1}) but violating (\ref{1qc}). Consider the membrane configuration in $T^{4}$ corresponding to string in $T^{3}$. Membrane should wrap one direction, for example, $x^{4}$ in $T^{4}$. Decompose the membrane worldvolume coordinate as $\xi^{i}=(\xi^{\hat{i}},\xi^{2})$, $\hat{i}=0,1$ and the target-space coordinate as $x^{\mu}=(x^{\hat{\mu}},x^{4})$, $\hat{\mu}=1,2,3$. One can let $\xi^{2}=x^{4}$ so that $\partial_{\hat{i}}x^{4}=0$, $\partial_{2}x^{4}=1$, and also suppose $\partial_{2}x^{\hat{\mu}}=0$. For simplicity, assume $g_{4\hat{\mu}}=b_{\hat{\mu}\hat{\nu}\hat{\rho}}=0$. One can prove that the transformation 
\begin{eqnarray}
\nonumber && 	\delta \partial_{\hat{i}}x^{\hat{\mu}}=C^{\hat{\rho}\hat{\mu}4}(-g_{\hat{\rho}\hat{\nu}}\sqrt{-\gamma}\gamma^{\hat{k}\hat{j}}\epsilon_{\hat{i}\hat{k}}\partial_{\hat{j}}x^{\hat{\nu}}+b_{\hat{\rho}\hat{\nu}4}\partial_{\hat{i}}x^{\hat{\nu}})
\\
 && 	\delta \partial_{\hat{k}}x^{4}=\delta \partial_{2}x^{\hat{\mu}}=\delta \partial_{2}x^{4}=0
\end{eqnarray}
is integrable on-shell. $j^{3\hat{\mu}\hat{\nu}}$ does not transform as in (\ref{1qc}) but (\ref{cf1}) is respected. So for such kind of the particular membrane configurations, the symmetry generated by $C^{\mu\nu4}$ can indeed be realized at the price of the violation of (\ref{1qc}). Compared with (\ref{4ew}), $C^{\mu\nu4}\sim B^{\mu\nu}$ generates part of the $SO(3,3)$ T-duality transformation for string in $T^{3}$.

Besides, there are several subtleties in the U-duality transformation of the membrane. Consider M theory on $T^{4}$ and then make a dimensional reduction along $x^{4}$ to get type IIA in $T^{3}$. The $SL(5,Z)$ U-duality also has its manifestation in type IIA. For example, M2 without wrapping $T^{4}$ becomes D2 without wrapping $T^{3}$, which is dual to D4 wrapping $12$ or $13$ or $23$, or equivalently, M5 wrapping $124$ or $134$ or $234$. In this case, M2 is U-dual to M5. M2 wrapping $23$ is also D2 wrapping $23$, which, after the T-duality transformation along $2$ becomes D1 winding $3$ with all winding modes along $2$ included, if the original M2 configuration is not translation invariant along the $x^{2}$ direction. A further T-duality transformation along $1$ gives D2 wrapping $13$, or equivalently, M2 wrapping $13$, which is translation invariant along $1$ but still with all winding modes along $2$ included. Obviously, the appearance of M5 and the infinite number of the winding modes after the U-duality transformation cannot be realized in the worldvolume transformation of the membrane.

It is well-known that matrix theory \cite{twsl}, which is supposed to be the non-perturbative description of M theory, is U-duality invariant \cite{twsla,twslb,twslc,twsld,twsle}. On the other hand, matrix theory can also arise as the discrete regularization of the membrane theory in lightcone gauge \cite{twsl1}, so U-duality transformation rule of matrix theory can also be translated as the U-duality transformation rule of membrane. Moreover, matrix theory description is also complete enough to incorporate the M5 brane as well as the winding modes.

In the following, we will consider the membrane worldvolume theory in lightcone gauge, which is equivalent to the matrix theory. We will discuss the U-duality transformation rule for matrix theory in $T^{n}\times R$, and then reinterpret it as the U-duality transformation rule of the membrane theory in $T^{n}\times R$. This is a little different from the approaches in \cite{dd1,ddd1}, where membrane is the instanton in $T^{n}$. As a result, some conclusions in \cite{dd1,ddd1} do not hold here. For example, membrane in $T^{3}$ is topological with the identity (\ref{ll}) holds, while the membrane in $T^{3}\times R$ is dynamical with the momentum density and the winding number density independent.

\section{A review of the matrix theory description of the membrane in $T^{n}\times R$}

Let us consider (the bosonic part of) the supermembrane action with the background fields $g_{\mu\nu}$ and $b_{\mu\nu\lambda}$, $\mu=0,1,\cdots,10$ \cite{asz12}.
\begin{equation}\label{ft}
	S=-\int  d^{3}\sigma \; \; [\sqrt{-\gamma}(\gamma^{\alpha\beta}\partial_{\alpha}x^{\mu}\partial_{\beta}x^{\nu}g_{\mu\nu}-1)+\partial_{0}x^{\mu}\{x^{\nu},x^{\lambda}\}b_{\mu\nu\lambda}]\;,
\end{equation}
where $\{x^{\nu},x^{\lambda}\}=\epsilon^{ab}\partial_{a}x^{\nu}\partial_{b}x^{\lambda}$, $\epsilon^{12}=1$, $\gamma_{\alpha\beta}=\partial_{\alpha}x^{\mu}\partial_{\beta}x^{\nu}g_{\mu\nu}$, $\alpha,\beta=0,1,2$, $a,b=1,2$. Suppose the membrane world-volume is of the form $\Sigma \times R$, where $\Sigma$ is a Riemann surface of the fixed topology and $R$ is the time direction $\sigma_{0}$. In lightcone gauge,
\begin{equation}
	x^{\pm}=\frac{x^{0}\pm x^{10}}{\sqrt{2}}\;,
\end{equation}
with the gauge fixing $\gamma_{0a}=0$, (\ref{ft}) is equivalent to  
\begin{equation}\label{fg6}
	S = \nu \int d^{3}\sigma \; \;(\frac{1}{2}D_{0}x^{i} D_{0}x^{j}  g_{ij} -\frac{1}{\nu} D_{0}x^{i} \{x^{j},x^{k}  \} b_{ijk}     - \frac{1}{\nu^{2}} \{x^{i},x^{j}  \} \{x^{k},x^{l}  \} g_{ik} g_{jl}  ) \;,
\end{equation}
where 
\begin{equation}
	D_{0}x^{i}=\partial_{0}x^{i}-\{\omega,x^{i}\}\;,
\end{equation}
$i=1,2,\cdots ,9$. In temporal gauge, we have $\omega=0$ together with the constraint 
\begin{equation}
	\{\partial_{0}x^{i},x_{i}\}=0\;.
\end{equation}
With $x^{+}=\sigma_{0}\equiv t$, the Hamiltonian is 
\begin{equation}\label{H}
	H = \nu \int  d^{2}\sigma \; \;(\frac{1}{2}\dot{x}^{i} \dot{x}^{j}  g_{ij} + \frac{1}{\nu^{2}} \{x^{i},x^{j}  \} \{x^{k},x^{l}  \} g_{ik} g_{jl}  ) \;,
\end{equation}
where $\dot{x}^{i}=\partial_{t}x^{i}=\partial_{0}x^{i}$. The conjugate momentum is
\begin{equation}
	p_{i}=g_{ij}\dot{x}^{j}-\frac{1}{\nu}\{x^{j},x^{k}\}b_{ijk}\;.
\end{equation}
The equation of motion is 
\begin{equation}\label{32}
	\ddot{x}^{i}=\frac{4}{\nu^{2}}\{\{x^{i},x^{j}\},x_{j}\}\;.
\end{equation}

With the matrix regularization \cite{twsl1}
\begin{equation}\label{35}
x(t,\sigma_{1},\sigma_{2})\rightarrow X(t)~,\;\;\;\;\{f,g\}\rightarrow -\frac{iN}{2}[F,G]~,\;\;\;\;\frac{1}{4\pi}\int d^{2}\sigma\; f=\frac{1}{N} tr\; F~,
\end{equation}
(\ref{fg6}) with $\omega=0$ becomes 
\begin{equation}\label{bv}
	S = 4\pi\kappa\;\int dt \;tr  \;(\frac{1}{2}\dot{X}^{i} \dot{X}^{j}  g_{ij} +\frac{i}{2\kappa} \dot{X}^{i} [X^{j},X^{k} ] b_{ijk}     + \frac{1}{4\kappa^{2}} [X^{i},X^{j}  ][X^{k},X^{l} ] g_{ik} g_{jl}  ) 
\end{equation}
with the constraint
\begin{equation}\label{65t}
	[\dot{X}^{i},X_{i}]=0\;,
\end{equation}
where $\nu/N=\kappa$. (\ref{bv}) is the matrix theory action on the background $g_{ij}$ and $b_{ijk}$ with $X^{i}$ the $N \times N$ matrix, $N=\infty$. The extension to the supersymmetric case is straightforward. (\ref{bv}) is also the low energy effective action for $N$ D0-branes.

When a particular transverse dimension $x^{\hat{i}}$ is compactified to $S^{1}$ with the radius $R^{\hat{i}}$, $X^{i}$ for $i=1,2,\cdots ,9$ should be replaced by the infinite block matrix $X^{i}_{mn}$ with constraints \cite{const1}
\begin{eqnarray}\label{re3}
X^i_{mn} & = & X^i_{(m-1)(n-1)}=X^i_{m-n}\;,\;\;\;\;\;  i \neq \hat{i} \nonumber\\
X^{\hat{i}}_{mn} & = & X^{\hat{i}}_{(m-1)(n-1)}=X^{\hat{i}}_{m-n}\; ,\;\;\;\;\;  m\neq n \\
X^{\hat{i}}_{nn} & = & 2 \pi R^{\hat{i}}  +X^{\hat{i}}_{(n-1)(n-1)}=2n \pi R^{\hat{i}}  +X^{\hat{i}}_{0}\; . \nonumber
\end{eqnarray}
$X^{\hat{i}}_{mn}$ can be explicitly written as
\begin{equation}
\left(\begin{array}{ccccccc}
\ddots& X^{\hat{i}}_{1} & X^{\hat{i}}_{2} & X^{\hat{i}}_{3} & \ddots \\
X^{\hat{i}}_{-1} & X^{\hat{i}}_{ 0} -2 \pi R^{\hat{i}}  & X^{\hat{i}}_{1} & X^{\hat{i}}_{2} & X^{\hat{i}}_{3} \\
X^{\hat{i}}_{-2} & X^{\hat{i}}_{-1}&X^{\hat{i}}_0 & X^{\hat{i}}_{1}&X^{\hat{i}}_{2}  \\
X^{\hat{i}}_{-3} &X^{\hat{i}}_{ -2} & X^{\hat{i}}_{-1} & X^{\hat{i}}_{0}  + 2 \pi R^{\hat{i}}  &X^{\hat{i}}_{1} \\
\ddots & X^{\hat{i}}_{ -3} & X^{\hat{i}}_{-2} & X^{\hat{i}}_{ -1} & \ddots
\end{array} \right)
\end{equation}
where $ X^{\hat{i}}_{0}$ is the original $X^{\hat{i}}$, $X^{\hat{i}}_{m-n}$ for $m \neq n$ are winding modes. If there are $n$ transverse dimensions $x^{\hat{i}}$ with $\hat{i}=1,2,\cdots ,n$ compactified to $S^{1}$, we can make the replacement (\ref{re3}) for $n$ directions successively.

After a T-duality transformation, D0-brane with one transverse dimension compactified with the radius $R^{\hat{i}}$ becomes D1-brane with one longitudinal dimension compactified with the radius $\tilde{R}^{\hat{i}}=\alpha'/R^{\hat{i}}=1/(2\pi R^{\hat{i}})$.
\begin{equation}
	X^{\hat{i}}=i \partial^{\hat{i}}+A^{\hat{i}}
\end{equation}
where 
\begin{equation}
	i \partial^{\hat{i}}=diag(\cdots, -4\pi R^{\hat{i}}, -2\pi R^{\hat{i}},0,2\pi R^{\hat{i}},4\pi R^{\hat{i}},\cdots).
\end{equation}
\begin{equation}
	A^{\hat{i}}_{mn}=A^{\hat{i}}_{m-n}=X^{\hat{i}}_{m-n}
\end{equation}
are the momentum modes of the gauge field on the dual circle
\begin{equation}
A^{\hat{i}}(\tilde{x})=\sum_{n}A^{\hat{i}}_{n}e^{in\tilde{x}/\tilde{R}^{\hat{i}}}.
\end{equation}
After the successive $n$ times T duality transformations, D0 branes with $n$ transverse dimension compactified to $R^{\hat{i}}$ becomes D$n$ branes with $n$ longitudinal dimensions compactified to $\tilde{R}^{\hat{i}}$, $\hat{i}=1,\cdots,n$. (\ref{bv}) becomes (the bosonic part of) the action of the $(n+1)$-dimensional SYM theory on the dual $\tilde{T}^{n}$ in temporal gauge with $A^{0}=0$. The winding modes on $T^{n}$ are converted into the momentum modes of the gauge fields on the dual $\tilde{T}^{n}$. In the following, we will consider two particular situations with $n=3$ and $n=4$.

\section{U-duality transformation of membrane on $T^{3}\times R$}

In this case, $x^{\hat{i}}$ with $\hat{i}=1,2,3$ are compactified to $S^{1}$. $\{i\}=\{\hat{i}\}\cup\{\bar{i}\}$, where $\bar{i}=4,6,\cdots,10$, $i=1,\cdots,4,6,\cdots,10$\footnote{Different from section 4, where $i=1,\cdots,9$, in section 5 and 6, we will assume $i=1,\cdots,4,6,\cdots,10$ for convenience.}. $R$ is identified with the time direction $t$, which, together with $x^{i}$, composes the $10d$ spacetime. $(g_{ij},b_{ijk})\equiv (g_{\hat{i}\hat{j}}, g_{\hat{i}\bar{j}}, g_{\bar{i}\bar{j}}, b_{\hat{i}\hat{j}\hat{k}}, b_{\hat{i}\hat{j}\bar{k}}, b_{\hat{i}\bar{j}\bar{k}}, b_{\bar{i}\bar{j}\bar{k}})$. After three T-duality transformations, (\ref{bv}) becomes (the bosonic part of) the action of the $\mathcal{N} = \textnormal{4}$ SYM theory on the dual $\tilde{T}^{3}$ in the temporal gauge with $A^{0}=0$. Let $I=0,1,2,3$ and set the six transverse scalar fields $X^{\bar{i}}=0$ for simplicity, equations of motion and the Bianchi identity for the YM fields on $\tilde{T}^{3}\times R$ are 
\begin{eqnarray}
  && D_{J}F^{IJ}=0\;,\label{syda}\\
 && \epsilon^{IJKL}D_{J}F_{KL}=0\;. \label{sdda}
\end{eqnarray}

The original membrane configuration $x^{\hat{i}}(t,\sigma_{1},\sigma_{2})$ corresponds to $X^{\hat{i}}(t)$, which is the zero mode $A^{\hat{i}}_{0}(t)$ of the YM fields on $\tilde{T}^{3}$. For the zero mode, 
\begin{eqnarray}
  && 	F^{\hat{i}\hat{j}}=-i[A^{\hat{i}},A^{\hat{j}}]=-i[X^{\hat{i}},X^{\hat{j}}]\;,\\
 && F^{\hat{i}0}=\dot{A}^{\hat{i}}=\dot{X}^{\hat{i}}\;.
\end{eqnarray}
(\ref{syda}) and (\ref{sdda}) reduce to 
\begin{equation}\label{123wsa}
[A_{\hat{i}},\dot{A}^{\hat{i}}]=[X_{\hat{i}},\dot{X}^{\hat{i}}]=0\;,\;\;\;\;\;\ddot{A}^{\hat{j}}+[A_{\hat{i}},[A^{\hat{i}},A^{\hat{j}}]]=\ddot{X}^{\hat{j}}+[X_{\hat{i}},[X^{\hat{i}},X^{\hat{j}}]]=0\;,
\end{equation}
which is the equation of motion in matrix theory together with the Gauss constraint, or equivalently, the equation of motion for membrane in lightcone gauge.

Wrapping number of the membrane is mapped as
\begin{equation}\label{szx}
W^{\hat{i}\hat{j}}=\frac{1}{A}	\int d^{2}\sigma\; \{	x^{\hat{i}},	x^{\hat{j}}\}=-i \;tr[A^{\hat{i}},A^{\hat{j}}]=tr F^{\hat{i}\hat{j}}\;,
\end{equation}
while the momentum becomes 
\begin{equation}\label{szxx}
	P_{\hat{k}}=\frac{1}{A}	\int d^{2}\sigma\; (g_{\hat{k}\hat{j}}\dot{x}^{\hat{j}}-\frac{1}{\nu}\{x^{\hat{i}},x^{\hat{j}}\}b_{\hat{i}\hat{j}\hat{k}})=tr (g_{\hat{k}\hat{j}}\dot{A}^{\hat{j}}-i[A^{\hat{i}},A^{\hat{j}}]b_{\hat{i}\hat{j}\hat{k}})=tr (g_{\hat{k}\hat{j}}F^{\hat{j}0}-F^{\hat{i}\hat{j}}b_{\hat{i}\hat{j}\hat{k}})\;.
\end{equation}
In (\ref{szx}) and (\ref{szxx}), $A^{\hat{i}}$ is the 0-mode of the gauge field. If $A^{\hat{i}}$ is the $N\times N$ matrix with $N$ finite, there will be $W^{\hat{i}\hat{j}}=0$. In matrix theory, the non-trivial wrapping number can be produced due to $N=\infty$.

For the membrane configuration in $T^{n} \times R$ with the topology of $T^{2} \times R$, one can make a mode expansion
\begin{equation}\label{dxc1}
x^{\hat{i}}(t,\sigma_{1},\sigma_{2})=a^{\hat{i}}\sigma_{1}+b^{\hat{i}}\sigma_{2}+c^{\hat{i}}t+\sum^{\infty}_{k^{1},k^{2}=-\infty} d^{\hat{i}}_{(k^{1},k^{2})}(t)e^{ik^{1}\sigma_{1}+ik^{2}\sigma_{2}}
\end{equation}
with $a^{\hat{i}}$ and $b^{\hat{i}}$ windings around $\sigma_{1}$ and $\sigma_{2}$ respectively, which are integers
and thus must be time independent. The wrapping number is calculated to be
\begin{equation}
W^{\hat{i}\hat{j}}=a^{\hat{i}}b^{\hat{j}}-a^{\hat{j}}b^{\hat{i}}\;,
\end{equation}
while the momentum is
\begin{equation}
	P_{\hat{k}}=g_{\hat{k}\hat{j}}c^{\hat{j}}-(a^{\hat{i}}b^{\hat{j}}-a^{\hat{j}}b^{\hat{i}})b_{\hat{i}\hat{j}\hat{k}}\;.
\end{equation}
$(\sigma_{1},\sigma_{2})$ are also subject to a $SL(2,Z)$ transformation, under which, $W^{\hat{i}\hat{j}}$ and $P_{\hat{k}}$ are invariant. With the replacement 
\begin{equation}
	\sigma_{1}\rightarrow Y_{1}\;,\;\;\;\;\;\;\;	\sigma_{2}\rightarrow Y_{2}\;,\;\;\;\;\;\;\;e^{ik^{1}\sigma_{1}+ik^{2}\sigma_{2}}\rightarrow Y^{(k^{1},k^{2})}\;,
\end{equation}
where $Y_{1}$, $Y_{2}$ and $Y^{(k^{1},k^{2})}$ are $N \times N$ matrices with $N=\infty$ and $[Y_{1},Y_{2}]=iI_{N\times N}$, we get the matrix configuration
\begin{equation}
	x^{\hat{i}}(t,\sigma_{1},\sigma_{2})\rightarrow X^{\hat{i}}(t)=a^{\hat{i}}Y_{1}+b^{\hat{i}}Y_{2}+c^{\hat{i}}t+\sum^{\infty}_{k^{1},k^{2}=-\infty} d^{\hat{i}}_{(k^{1},k^{2})}(t)Y^{(k^{1},k^{2})}
\end{equation}
with wrapping taken into account.

$\mathcal{N} = \textnormal{4}$ SYM theory is expected to have the $SL(3,Z)\times SL(2,Z)$ duality symmetry. For $A^{\hat{i}}(x^{\hat{j}})$ on $T^{3}$, the $SL(3,Z)$ transformation acts on both the coordinate $x^{\hat{j}}$ and the index $\hat{i}$. For the 0-mode $A^{\hat{i}}_{0}=X^{\hat{i}}\sim x^{\hat{i}}(\sigma_{1},\sigma_{2})$, $SL(3,Z)$ only acts on the index and thus will manifest as a global symmetry of the membrane worldvolume, i.e., $x^{\hat{i}}(\sigma_{1},\sigma_{2}) \rightarrow x'^{\hat{i}}(\sigma_{1},\sigma_{2}) = \Lambda^{i}_{j}x^{\hat{j}}(\sigma_{1},\sigma_{2})$ for $\Lambda\in SL(3,Z)$. Let us define
\begin{equation}
	\tau=b+i|\hat{g}|^{1/2},
\end{equation}
where $b_{\hat{i}\hat{j}\hat{k}}=\epsilon_{\hat{i}\hat{j}\hat{k}}b$, $\hat{g}=\det g_{\hat{i}\hat{j}}$. Under the $SL(2,Z)$ S-duality transformation, $\tau$ transforms as 
\begin{equation}
	\tau\rightarrow\frac{a \tau+b}{c\tau +d} \;\;\;\;\;\;\;\;\;ad-bc=1,\;\;\;\;\;a,b,c,d\in Z.
\end{equation}

The $SL(2,Z)$ transformation is generated by the T element 
\begin{equation}
	\tau\rightarrow \tau +1
\end{equation}
and the S element 
\begin{equation}
	\tau \rightarrow -\frac{1}{\tau}\;.
\end{equation}
The T transformation makes $F_{KL}$ invariant, but the S transformation is quite nontrivial. When $N=1$, (\ref{syda}) and (\ref{sdda}) become 
\begin{eqnarray}
 && \partial_{J}F^{IJ}=0\;,\label{ss1}\\
 && \epsilon^{IJKL}\partial_{J}F_{KL}=0\;.\label{ss}
\end{eqnarray}
S transformation is simply 
\begin{equation}\label{gg1}
	F_{IJ}\rightarrow -\frac{1}{2}\epsilon_{IJKL}F^{KL}\;.
\end{equation}
For $N>1$, (\ref{syda}) and (\ref{sdda}) are
\begin{eqnarray}
 && \partial_{J}F^{IJ}-i [A_{J},F^{IJ}]=0\;,\label{asz1}\\
 && \epsilon^{IJKL}\partial_{J}F_{KL}-i\epsilon^{IJKL} [A_{J},F_{KL}]=0\;,\label{asz}
\end{eqnarray}
for which (\ref{gg1}) does not necessarily apply anymore.

Nevertheless, there are some special non-abelian field configurations with (\ref{gg1}) valid. One such example is when $F_{IJ}= \tilde{F}_{IJ}=-\epsilon_{IJKL}F^{KL}/2$ and then $A_{I}=\tilde{A}_{I}$. Another example is 
\begin{equation}\label{1az}
		A^{\hat{i}}(t)=a^{\hat{i}}Y_{1}+b^{\hat{i}}Y_{2}+c^{\hat{i}}t\;, 
\end{equation}
which trivially satisfies the equations of motion.
\begin{equation}
	F^{\hat{i}\hat{j}}=(a^{\hat{i}}b^{\hat{j}}-a^{\hat{j}}b^{\hat{i}})I_{N\times N}\;,\;\;\;\;\;\;\;\;\;	F^{0\hat{i}}=c^{\hat{i}}I_{N\times N}\;.
\end{equation}
The S-dual gauge field can be taken to be
\begin{equation}\label{1azz}
\tilde{A}^{\hat{i}}(t)=\tilde{a}^{\hat{i}}Y_{1}+\tilde{b}^{\hat{i}}Y_{2}+\tilde{c}^{\hat{i}}t\;
\end{equation}
with 
\begin{equation}\label{k5}
\tilde{c}_{\hat{i}}=-	\epsilon_{\hat{i}\hat{j}\hat{k}}a^{\hat{j}}b^{\hat{k}}\;,\;\;\;\;\;\;\;\;	\epsilon_{\hat{i}\hat{j}\hat{k}}\tilde{a}^{\hat{i}}\tilde{b}^{\hat{j}}=c_{\hat{k}}\;
\end{equation}
so that the field strength transforms as 
\begin{equation}
	F_{0\hat{i}}\rightarrow -\frac{1}{2} \epsilon_{\hat{i}\hat{j}\hat{k}} F^{\hat{j}\hat{k}}  \;,\;\;\;\;\;\;\;\;   \frac{1}{2}\epsilon^{\hat{i}\hat{j}\hat{k}} F_{\hat{i}\hat{j}}\rightarrow   F^{\hat{k}0}\;.
\end{equation}
(\ref{1az}) and (\ref{1azz}) can be mapped into the membrane configuration 
\begin{equation}
	x^{\hat{i}}(t,\sigma_{1},\sigma_{2})=a^{\hat{i}}\sigma_{1}+b^{\hat{i}}\sigma_{2}+c^{\hat{i}}t
\end{equation}
and its S-dual 
\begin{equation}
\tilde{x}^{\hat{i}}(t,\sigma_{1},\sigma_{2})=\tilde{a}^{\hat{i}}\sigma_{1}+\tilde{b}^{\hat{i}}\sigma_{2}+\tilde{c}^{\hat{i}}t\;.
\end{equation}
Both of them are 1/2 BPS configurations and the transformation law (\ref{k5}) can be verified by considering the T-duality transformation of $F1$ in type IIA theory compactified on $T^{2}$.

As the third example, consider the gauge fields 
\begin{eqnarray}\label{1q2z}
 \nonumber && A_{1}(t)= c_{1}t+b_{1}Y_{2}+\sum^{\infty}_{k^{2}=-\infty} d^{L}_{1(0,k^{2})}e^{ik^{2}(Y_{2}+t)}+ d^{R}_{1(0,k^{2})}e^{ik^{2}(Y_{2}-t)}\;,\\ \nonumber
 &&  A_{2}(t)= c_{2}t+b_{2}Y_{2}+\sum^{\infty}_{k^{2}=-\infty} d^{L}_{2(0,k^{2})}e^{ik^{2}(Y_{2}+t)}+ d^{R}_{2(0,k^{2})}e^{ik^{2}(Y_{2}-t)}\;,\\
 &&A_{3}(t)=Y_{1}\;,
\end{eqnarray}
which is a solution for the equations of motion (\ref{123wsa}). The dual field $(\tilde{A}_{1},\tilde{A}_{2},\tilde{A}_{3})$ can be obtained via the replacement 
\begin{equation}\label{Udd}
\tilde{b}_{\alpha}=\epsilon_{\alpha\beta}c_{\beta}\;\;\;\;\;\;\;\;\;\tilde{c}_{\alpha}=\epsilon_{\alpha\beta}b_{\beta}\;\;\;\;\;\;\;\;\;\tilde{d}^{L}_{\alpha(0,k^{2})}=-\epsilon_{\alpha\beta}d^{L}_{\beta(0,k^{2})}\;\;\;\;\;\;\;\;\;\tilde{d}^{R}_{\alpha(0,k^{2})}=\epsilon_{\alpha\beta}d^{R}_{\beta(0,k^{2})}
\end{equation}
and is again a solution for the equations of motion, $\alpha,\beta=1,2$. The hodge dual relation $\tilde{F}_{IJ}= -\epsilon_{IJKL}F^{KL}/2$ is satisfied. In fact, (\ref{1q2z}) can be mapped to the membrane configuration 
\begin{eqnarray}
 \nonumber && x_{1}= c_{1}t+b_{1}\sigma_{2}+\sum^{\infty}_{k^{2}=-\infty} d^{L}_{1(0,k^{2})}e^{ik^{2}(\sigma_{2}+t)}+ d^{R}_{1(0,k^{2})}e^{ik^{2}(\sigma_{2}-t)}\;,\\ \nonumber
 &&  x_{2}= c_{2}t+b_{2}\sigma_{2}+\sum^{\infty}_{k^{2}=-\infty} d^{L}_{2(0,k^{2})}e^{ik^{2}(\sigma_{2}+t)}+ d^{R}_{2(0,k^{2})}e^{ik^{2}(\sigma_{2}-t)}\;,\\
 &&x_{3}=\sigma_{1}\;,
\end{eqnarray}
which is also the string in type IIA theory compactified on $T^{2}$ with $x^{3}$ the M theory direction. U-duality transformation (\ref{Udd}) can be realized via two successive $O(2,2;Z)$ T duality transformations together with a $SL(2,Z)$ transformation on $T^{2}$.

For the generic on-shell non-abelian gauge field $A_{I}$, the S-dual field strength cannot satisfy $\tilde{F}_{IJ}= -\epsilon_{IJKL}F^{KL}/2$, since the related gauge field $\tilde{A}_{I}$ may not exist. The key point for the S transformation is that the equations of motion for $F^{IJ}$ can be reinterpreted as the integrable condition for the dual field strength $\tilde{F}^{IJ}$, and vice versa. Based on the loop space formulation, \cite{1g, 7hg} give a prescription to get the S-dual for the generic non-abelian gauge field. Consider the loops passing through a fixed reference point $\xi_{0}$
\begin{equation}
	C:\;\;\;\;\;\;\{\xi^{I}(s): s=0\rightarrow 2\pi,\xi(0)=\xi(2\pi)=\xi_{0}\}\;.
\end{equation}
For each loop, one may define a path-ordered phase factors (Wilson loops)
\begin{equation}
	\Phi[\xi]=P_{s} \exp ig \int^{2\pi}_{0} ds A_{I}(\xi(s))\dot{\xi}^{I}(s)\;.
\end{equation}
The derivatives in loop space can be defined as 
\begin{equation}
	\delta_{I}(s)\Psi[\xi]\equiv \lim_{\Delta\rightarrow 0}\frac{1}{\Delta}\{\Psi[\xi']-\Psi[\xi]\}\;,
\end{equation}
with
\begin{equation}
	\xi'^{J}(s')=\xi^{J}(s')+\Delta\delta^{J}_{I}\delta(s-s')\;. 
\end{equation}
A new variable $E_{I}[\xi|s]$ can be introduced as follows
\begin{equation}
	E_{I}[\xi|s]=\Phi_{\xi}(s,0)\frac{i}{g}\Phi^{-1}[\xi]\delta_{I}(s)\Phi[\xi]\Phi^{-1}_{\xi}(s,0)\;,
\end{equation}
where
\begin{equation}
	\Phi_{\xi}(s_{2},s_{1})=P_{s} \exp ig \int^{s_{2}}_{s_{1}} ds A_{I}(\xi(s))\dot{\xi}^{I}(s)\;.
\end{equation}
In order to guarantee the existence of $A_{I}(x)$, from which, $E_{I}[\xi|s]$ can be derived, $E_{I}[\xi|s]$ should satisfy the integrable condition 
\begin{equation}\label{1o}
	\delta_{I}(s)E_{J}[\xi|s]-	\delta_{J}(s)E_{I}[\xi|s]=0\;	.
\end{equation}
Besides, to make $A_{I}(x)$ satisfy the Yang-Mills equation (\ref{syda}), $E_{I}[\xi|s]$ should also satisfy 
\begin{equation}\label{11o}
		\delta^{I}(s)E_{I}[\xi|s]=0\;.
\end{equation}
(\ref{1o}) and (\ref{11o}) are the integrable condition and the equation of motion for the loop space variable $E_{I}[\xi|s]$. The S-dual $\tilde{E}_{I}$ is defined as 
\begin{equation}\label{fg1}
\omega^{-1}(\eta(t))	\tilde{E}_{I}[\eta|t]\omega(\eta(t))=-\frac{2}{\bar{N}}\epsilon_{IJKL}\dot{\eta}^{J}(t)\int\delta\xi ds \; E^{K}[\xi|s]\dot{\xi}^{L}(s)\dot{\xi}^{-2}(s)\delta(\xi(s)-\eta(t))\;,
\end{equation}
or more concretely, 
\begin{equation}\label{fg11}
\omega^{-1}(x)	\tilde{F}_{IJ}(x)\omega(x)=-\frac{2}{\bar{N}}\epsilon_{IJKL}\int\delta\xi ds \; E^{K}[\xi|s]\dot{\xi}^{L}(s)\dot{\xi}^{-2}(s)\delta(x-\xi(s))\;.
\end{equation}
As is required, the dual transformation is reversible apart from a sign, i.e., $\tilde{\tilde{E}}=-E$. For $N=1$, (\ref{fg11}) gives the abelian S-dual $\tilde{F}_{IJ}=-\frac{1}{2}\epsilon_{IJKL}F^{KL}$. From (\ref{fg1}), one can prove 
\begin{equation}
			\delta^{I}(s)E_{I}[\xi|s]=0 \;\;\; \Leftrightarrow \;\;\; \delta_{I}(t)\tilde{E}_{J}[\eta|t]-	\delta_{J}(t)\tilde{E}_{I}[\eta|t]=0\;	,
\end{equation}
\begin{equation}
		\delta_{I}(s)E_{J}[\xi|s]-	\delta_{J}(s)E_{I}[\xi|s]=0	\;\;\; \Leftrightarrow \;\;\; \delta^{I}(t)\tilde{E}_{I}[\eta|t]=0\;,
\end{equation}
which is the non-abelian extension of
\begin{equation}
		\partial_{J}F^{IJ}=0 \;\;\; \Leftrightarrow \;\;\; \epsilon^{IJKL}\partial_{J}\tilde{F}_{KL}=0\;	,
\end{equation}
\begin{equation}
	\epsilon^{IJKL}\partial_{J}F_{KL}=0	\;\;\; \Leftrightarrow \;\;\; 	\partial_{J}\tilde{F}^{IJ}=0\;.
\end{equation}
Taking the trace on both sides of (\ref{fg11}), we get 
\begin{equation}
{\rm tr}\tilde{F}_{IJ}=-\frac{1}{2}\epsilon_{IJKL}{\rm tr} F^{KL}\;.
\end{equation}
Thus, although the relation between $ F_{IJ}$ and its S-dual $\tilde{F}_{IJ}$ may be complicated, their traces respect the simple hodge dual relation. For the 0-mode, ${\rm tr} F^{0\hat{i}}$ and $\frac{1}{2}\epsilon_{\hat{i}\hat{j}\hat{k}}{\rm tr} F^{\hat{j}\hat{k}}$ are the membrane momentum and wrapping number respectively. If the original field $A$ and dual field $\tilde{A}$ are both 0-modes, since
\begin{equation}\label{sd}
{\rm tr} \tilde{F}_{0\hat{i}}=-\frac{1}{2}\epsilon_{\hat{i}\hat{j}\hat{k}}{\rm tr} F^{\hat{j}\hat{k}}\;, \;\;\;\;\;\;\;-\frac{1}{2}\epsilon^{\hat{i}\hat{j}\hat{k}}{\rm tr}\tilde{F}_{\hat{j}\hat{k}}={\rm tr} F^{0\hat{i}}\;,
\end{equation}
S transformation exchanges the momentum and wrapping number as is required. On the other hand, (\ref{sd}) does not hold without the trace, so S transformation does not exchange the momentum density and winding number density.

$\mathcal{N} = \textnormal{4}$ SYM theory can be taken as the $6d$ $(2,0)$ theory reduced along $4$ and $5$ directions. $A_{I}\equiv B_{I4}$, $F_{IJ}\equiv H_{IJ4}$. If we can find $H_{IJ5}\equiv\tilde{F}_{IJ}$, then $F_{IJ}$ and $\tilde{F}_{IJ}$ form the $SL(2,Z)$ doublet for $T^{2}\sim x^{4}\times x^{5}$. In the abelian case, the $3$-form strength in the $6d$ $(2, 0)$ theory is constrained by the self-dual relation
\begin{equation}\label{1q2w1}
	H_{\alpha\beta\gamma}=\frac{1}{6}\epsilon_{\alpha\beta\gamma\mu\nu\lambda}H^{\mu\nu\lambda}\;.
\end{equation}
$H_{IJ5}=\epsilon_{IJ5KL4}H^{KL4}/2$, or equivalently, $\tilde{F}_{IJ}=-\epsilon_{IJKL}F^{KL}/2$. In the non-abelian case, the loop space formulation may give the dual $\tilde{F}_{IJ}$, which could be taken as $H_{IJ5}$ in $6d$. In this sense, \cite{1g, 7hg} also give a generalization of the self-dual relation (\ref{1q2w1}) to the non-abelian case when the $6d$ theory is reduced on $T^{2}$.

It is the 0-mode $A_{I}(t)$ that is mapped to the membrane configuration. It remains to see whether the $SL(2,Z)\times SL(3,Z)$ transformation for $\mathcal{N} = \textnormal{4}$ SYM theory could map $A_{I}(t)$ to $A'_{I}(t)$ which is also constant in space. For the $SL(3,Z)$ transformation, the answer is obviously yes, but for the $SL(2,Z)$ transformation, there are counter-examples in $U(1)$ case. Consider the on-shell gauge field 
\begin{equation}
	A_{0}(t)=0\;,\;\;\;\;\;\;	A_{\hat{i}}(t)=K_{\hat{i}}t\;,
\end{equation}
\begin{equation}
B_{\hat{i}}=\frac{1}{2}	\epsilon_{\hat{i}\hat{j}\hat{k}}F^{\hat{j}\hat{k}}=0\;,\;\;\;\;\;\;	E_{\hat{i}}=F_{0\hat{i}}=K_{\hat{i}}\;.
\end{equation}
The dual field strength is
\begin{equation}
B_{\hat{i}}\rightarrow 	E_{\hat{i}}=K_{\hat{i}}	\;,\;\;\;\;\;\;	E_{\hat{i}}\rightarrow-B_{\hat{i}}=0\;,
\end{equation}
for which, the related $	A_{\hat{i}}(x,t)$ cannot be constant in space. In fact, $S$-transformation in $6d$ theory just makes 
\begin{equation}
	B_{4\hat{i}}(x_{\hat{k}},t)\rightarrow B_{5\hat{i}}(x_{\hat{k}},t)\;,\;\;\;\;\;\;\;\;\;B_{5\hat{i}}(x_{\hat{k}},t)\rightarrow -B_{4\hat{i}}(x_{\hat{k}},t)\;.
\end{equation}
$B_{4\hat{i}}$ and $B_{5\hat{i}}$ are not independent but are related via the self-duality relation. For $B_{4\hat{i}}/B_{5\hat{i}}$ that is constant in space, there is no reason to expect that $B_{5\hat{i}}/B_{4\hat{i}}$ is also constant. Especially, in the $U(1)$ case, the self-duality relation requires 
\begin{equation}
		H_{05\hat{i}}=\partial_{0} B_{5\hat{i}}+\partial_{5}B_{\hat{i}0}+\partial_{\hat{i}}B_{05}=\frac{1}{2}\epsilon_{05\hat{i}\hat{j}\hat{k}4 }H^{\hat{j}\hat{k}4}=\frac{1}{2}\epsilon_{05\hat{i}\hat{j}\hat{k}4 }(\partial_{\hat{j}} B_{\hat{k}4}+\partial_{\hat{k}}B_{4\hat{j}}+\partial_{4}B_{\hat{j}\hat{k}})\;,
\end{equation}
so if $B_{5\hat{i}}(t)$ is time-dependent, $B_{4\hat{i}}$ must be space-dependent. In non-abelian case, we have examples like (\ref{1az}) and (\ref{1q2z}), for which, $B_{4\hat{i}}$ and $B_{5\hat{i}}$ can both be constant in space, but generically, it is still expected that the $SL(2,Z)$ transformation do not always map the $0$-mode into the $0$-mode. According to (\ref{fg11}), starting from $A_{I}(t)$ that is constant in space, $F_{IJ}$ together with $\tilde{F}_{IJ}$ are also constant. Nevertheless, there may not be the constant $\tilde{A}_{I}$ with 
\begin{equation}
\tilde{F}_{\hat{i}\hat{j}}=-i[\tilde{A}_{\hat{i}},\tilde{A}_{\hat{j}}]\;,\;\;\;\;\;\;\tilde{F}_{0\hat{i}}=\dot{\tilde{A}}_{\hat{i}}\;.
\end{equation}

Membrane on $T^{n}$ contains more degrees of freedom than that in non-compact space, coming from the winding modes on $T^{n}$. Without the winding modes included, the membrane configuration corresponds to the constant gauge field of the SYM theory in $\tilde{T}^{n}$. With the winding degrees of freedom added, membrane is then related to the generic gauge field in $\tilde{T}^{n}$. When $n=3$, there is a $SL(2,Z)\times SL(3,Z)$ transformation, mapping the given $A_{I}(x,t)$ to $A'_{I}(x,t)$. If the 0-mode subspace is invariant under the U-duality transformation, i.e. $A_{I}(t) \rightarrow A'_{I}(t)$, there will be a well-defined U-duality transformation for membrane without the need of involving the winding modes. However, this is not quite likely to be the case and the U-duality transformation can only be complete with the winding modes also taken into account.

Finally, let us consider the complete picture for membrane living in $T^{3}\times R^{8}$, which has been discussed in \cite{ddd1,a,b,c}. In this case, membrane is not topological and it was found that only a two-dimensional Heisenberg subgroup of $SL(2)$ could be realized \cite{ddd1,b,c}. In lightcone gauge, the worldvolume theory of membrane in $T^{3}\times R^{8}$ (with winding mode added) is equivalent to $\mathcal{N}=4$ SYM theory in $\tilde{T}^{3}\times R^{6}\times R$, where the last $R$ represents the time dimension $t$. For simplicity, the nonvanishing background fields in $T^{3}\times R^{8}$ are taken to be $(g_{+-}=1, g_{\hat{i}\hat{j}},g_{\bar{i}\bar{j}}, b_{\hat{i}\hat{j}\hat{k}}, b_{\hat{i}\hat{j}\bar{k}}, b_{\hat{i}\bar{j}\bar{k}}, b_{\bar{i}\bar{j}\bar{k}})$, where $\hat{i}=1,2,3$, $\bar{i}=4,6,7,8,9,10$. All background fields depend on the coordinates $x^{\bar{i}}$ only. The nonvanishing dual background on $\tilde{T}^{3}\times R^{6}\times R$ is $(g_{tt}=-1, \tilde{\Phi}=-\frac{1}{2}\log |\hat{g}|,\tilde{g}_{\hat{i}\hat{j}},g_{\bar{i}\bar{j}}, b, b_{\hat{k}\bar{k}}, b_{\hat{j}\hat{k}\bar{j}\bar{k}}, b_{\hat{i}\hat{j}\hat{k}\bar{i}\bar{j}\bar{k}})$ with $\tilde{\Phi}$ the dilaton. Both the $SL(2,Z) \times SL(3,Z)$ U duality symmetry of the $11d$ supergravity on $T^{3}\times R^{8}$ and the $SL(2,Z) \times SL(3,Z)$ duality of type IIB supergravity on $\tilde{T}^{3}\times R^{6}\times R$ are well established. Moreover, $D3$ coupling with the type IIB supergravity is S duality invariant, so $\mathcal{N}=4$ SYM theory in type IIB background should also be $SL(2,Z)$ invariant.

The situation becomes complicated since the background fields may have the dependence on the six transverse coordinates $x^{\bar{i}}$, or in SYM theory, the six scalar fields $X^{\bar{i}}$. Multiple D branes in curved background was discussed in \cite{f}. For the term like $Tr[g_{\bar{i}\bar{j}}(X^{\bar{k}})(D^{\hat{i}}X^{\bar{i}}D_{\hat{i}}X^{\bar{j}})]$ with $g_{\bar{i}\bar{j}}$ depending on $x^{\bar{k}}$, one can make a Taylor expansion 
\begin{equation}
Tr[g_{\bar{i}\bar{j}}(X^{\bar{k}})(D^{\hat{i}}X^{\bar{i}}D_{\hat{i}}X^{\bar{j}})]=\sum^{\infty}_{n=0}\frac{1}{n!}\partial_{\bar{k}_{1}} \cdots \partial_{\bar{k}_{n}} g_{\bar{i}\bar{j}}(0)STr[D^{\hat{i}}X^{\bar{i}}D_{\hat{i}}X^{\bar{j}}X^{\bar{k}_{1}}\cdots X^{\bar{k}_{n}}]\; ,
\end{equation}
giving rise to an infinite number of the higher dimensional operators in the Lagrangian. We will not consider this generic situation here, since according to the counterexample given in \cite{ddd1}, even for the constant background, only a two-parameter subgroup of $SL(2)$ can be realized in $T^{3}\times R^{8}$.

We will focus on the simplest situation: $\mathcal{N}=4$ SYM theory in $10d$ flat spacetime with the constant axion-dilaton field $\tau=b+ie^{-\tilde{\Phi}}$. The remaining problem is to find a S transformation for SYM fields, extending the S transformation (\ref{fg11}) for YM fields. In \cite{k}, with the $4d$ spacetime extended to the $\mathcal{N}=1$ superspace, the loop space formulation and the S transformation for YM theory is extended to $\mathcal{N}=1$ SYM theory. The further extension to the $\mathcal{N}=4$ supersymmetrty is straightforward. However, just as the YM case, nothing could guarantee that the zero mode of the $\mathcal{N}=4$ SYM field would necessarily be mapped into the zero mode under the S transformation. So the $SL(2,Z)$ duality cannot always be realized at the membrane worldvolume level, in agreement with \cite{ddd1,b,c}. Nevertheless, with the winding modes on $T^{3}$ taken into account, membrane worldvolume theory on $T^{3}\times R^{8}$ is equivalent to the $\mathcal{N}=4$ SYM theory on $\tilde{T}^{3}\times R^{7}$, for which, the $SL(2,Z)\times SL(3,Z)$ symmetry is definite.

\section{U-duality transformation for membrane on $T^{4}\times R$}

The situation for membrane living in $T^{4}\times R$, wrapping the 2-cycles in $T^{4}$ is similar. Let $\hat{i}=1,2,3,4$ represent $4$ directions in $T^{4}$, $\bar{i}=6,7,\cdots,10$ represent $5$ uncompactified directions, $t\equiv x_{0}$ represent the time dimension. After four times of T-duality transformations, (\ref{bv}) becomes (the bosonic part of) the $\mathcal{N} = \textnormal{2}$ SYM theory on the dual $\tilde{T}^{4}$ in temporal gauge $A^{0}=0$ with the Gauss constraint 
\begin{equation}
	D_{\hat{i}}F^{\hat{i}0}=0\;.
\end{equation}
The $T^{4}$ independent background fields are denoted as $g_{\hat{i}\hat{j}}$ and $b_{\hat{i}\hat{j}\hat{k}}$, which will become $\tilde{g}_{\hat{i}\hat{j}}$ and $\tilde{b}_{\hat{i}}$ on the dual $\tilde{T}^{4}$. $\tilde{g}=g^{-1}$, $\tilde{b}_{\hat{i}}=\frac{1}{6}g_{\hat{i}\hat{j}}\epsilon^{\hat{j}\hat{k}\hat{l}\hat{m}}b_{\hat{k}\hat{l}\hat{m}}$. In type IIA theory, $\tilde{g}_{\hat{i}\hat{j}}$ and $\tilde{b}_{\hat{i}}$ altogether form a $5 \times 5$ metric $\tilde{G}_{\tilde{i}\tilde{j}}$ with the 5th direction the M theory dimension, $\tilde{i}=1,2,3,4,5$. 
\begin{equation}    
\tilde{G}_{\tilde{i}\tilde{j}}=\phi^{-\frac{2}{3}}\left(                
 \begin{array}{cc}   
\tilde{g}_{\hat{i}\hat{j}}+\phi^{2}\tilde{b}_{\hat{i}}\tilde{b}_{\hat{j}} & \phi^{2}\tilde{b}_{\hat{i}} \\ 
  \phi^{2}\tilde{b}_{\hat{j}}   & \phi^{2}\\  
 \end{array}
\right)  .               
\end{equation}
Under the $SL(5,Z)$ transformation, $\tilde{G}_{\tilde{i}\tilde{j}}$ transforms as $\tilde{G}\rightarrow U \tilde{G}U^{-1}$, $U \in SL(5,Z)$.

Let $I=0,1,2,3, 4$ and set $X^{\bar{i}}=0$ for simplicity, the equations of motion and the Bianchi identity for YM field on $\tilde{T}^{4}$ are 
\begin{eqnarray}
  && D_{J}F^{IJ}=0\;,\label{s1y}\\
 && \epsilon^{IJKLM}D_{K}F_{LM}=0\;. \label{s2}
\end{eqnarray}
For 0-mode, (\ref{s1y}) and (\ref{s2}) reduce to 
\begin{equation}\label{456}
[A_{\hat{i}},\dot{A}^{\hat{i}}]=[X_{\hat{i}},\dot{X}^{\hat{i}}]=0\;,\;\;\;\;\;\ddot{A}^{\hat{j}}+[A_{\hat{i}},[A^{\hat{i}},A^{\hat{j}}]]=\ddot{X}^{\hat{j}}+[X_{\hat{i}},[X^{\hat{i}},X^{\hat{j}}]]=0\;,
\end{equation}
which are the equations of motion in matrix theory together with the Gauss constraint, or the equations of motion for membrane in lightcone gauge. Membrane wrapping number is 
\begin{equation}
W^{\hat{i}\hat{j}}=-i \;tr[A^{\hat{i}},A^{\hat{j}}]=tr F^{\hat{i}\hat{j}}\;,
\end{equation}
while the momentum is
\begin{equation}
	P_{\hat{k}}=tr (g_{\hat{k}\hat{j}}\dot{A}^{\hat{j}}-i[A^{\hat{i}},A^{\hat{j}}]b_{\hat{i}\hat{j}\hat{k}})=tr (g_{\hat{k}\hat{j}}F^{\hat{j}0}-F^{\hat{i}\hat{j}}b_{\hat{i}\hat{j}\hat{k}})\;.
\end{equation}

The discussion is parallel to the $T^{3}$ situation. $\mathcal{N} = \textnormal{2}$ SYM theory is the $6d$ $(2, 0)$ theory reduced along $x_{5}$. For simplicity, let $b_{\hat{i}\hat{j}\hat{k}}=0$, $g_{\hat{k}\hat{j}}=\delta_{\hat{k}\hat{j}}$. $A_{I}\equiv B_{I5}$, $F_{IJ}\equiv H_{IJ5}$. $	H_{0\tilde{i}\tilde{j}}\equiv (H_{0\hat{i}5}, H_{0\hat{i}\hat{j}})$ forms a $5\times5$ antisymmetric matrix. In the abelian case, according to the self-duality relation (\ref{1q2w1}), $H_{0\hat{i}\hat{j}}$ is determined from $H_{\hat{i}\hat{j}5}$ via $H_{0\hat{i}\hat{j}}=\epsilon_{0\hat{i}\hat{j}\hat{k}\hat{l}5}H^{\hat{k}\hat{l}5}/2$, or equivalently, 
\begin{equation}\label{1031}
	H_{0\hat{i}\hat{j}}\equiv \tilde{F}_{\hat{i}\hat{j}}=\epsilon_{\hat{i}\hat{j}\hat{k}\hat{l}}F^{\hat{k}\hat{l}}/2\;.
\end{equation}
In the non-abelian case, $H_{0\hat{i}\hat{j}}$ and $H_{\hat{i}\hat{j}5}$ may still be related by some self-duality relation, whose form is unknown at present. One possibility is that we can still use the prescription in \cite{1g,7hg} but with the indices restricted to the $4d$ space
\begin{equation}\label{103}
 \omega^{-1}(x)H_{0\hat{i}\hat{j}} \omega(x)\equiv       \omega^{-1}(x)	\tilde{F}_{\hat{i}\hat{j}}(x)\omega(x)=\frac{2}{\bar{N}}\epsilon_{\hat{i}\hat{j}\hat{k}\hat{l}}\int\delta\xi ds \; E^{\hat{k}}[\xi|s]\dot{\xi}^{\hat{l}}(s)\dot{\xi}^{-2}(s)\delta(x-\xi(s))\;.
\end{equation}
(\ref{103}) reduces to (\ref{1031}) in the abelian case as is required. Recall that (\ref{fg11}) is supposed to determine $H_{5ij}/H_{4ij}$ from $H_{4ij}/H_{5ij}$ when the $6d$ theory is translation invariant along $x_{4}$ and $x_{5}$, $i=0,1,2,3$. The similar relation (\ref{103}) may be able to determine $H_{0\hat{i}\hat{j}}$ from $H_{5\hat{i}\hat{j}}$ when the $6d$ theory is translation invariant along $x_{0}$ and $x_{5}$, $\hat{i}=1,2,3,4$. So, (\ref{103}) is valid at least for the time-independent system. For the 0-mode, $tr H_{0\hat{i}5}=P_{\hat{i}}$ and $tr H^{\hat{i}\hat{j}5} =W^{\hat{i}\hat{j}}$ correspond to the membrane momentum and the wrapping number respectively. $H_{0\hat{i}\hat{j}}$ determined from (\ref{103}) satisfies the relation $tr H_{0\hat{i}\hat{j}}=\epsilon_{\hat{i}\hat{j}\hat{k}\hat{l}}tr H^{\hat{k}\hat{l}5}/2$. So if (\ref{103}) is valid, $tr	H_{0\tilde{i}\tilde{j}}\equiv (P_{\hat{i}}, \epsilon_{\hat{i}\hat{j}\hat{k}\hat{l}}W^{\hat{k}\hat{l}}/2)$ forms a $5\times5$ antisymmetric matrix.

With $H_{0\hat{i}\hat{j}}=H_{0\hat{i}\hat{j}}(H^{\hat{k}\hat{l}5})$ obtained, $SL(5,Z)$ transformation is realized as  
\begin{equation}\label{sx3s}
	H_{0\tilde{i}\tilde{j}}(x_{\tilde{k}},t)\rightarrow U^{\tilde{l}}_{\tilde{i}}U^{\tilde{m}}_{\tilde{j}} H_{0\tilde{l}\tilde{m}}(U^{\tilde{n}}_{\tilde{k}}x_{\tilde{n}},t)\;.
\end{equation}
However, for the $5d$ SYM theory, (\ref{sx3s}) is not well-defined due to a subtlety in $x_{5}$ direction: fields are translation invariant along $x_{5}$ but not necessarily so in $1234$ space. In fact, gauge field configurations in $\mathcal{N} = \textnormal{2}$ SYM theory are classified by the instanton number
\begin{equation}
	n=-\frac{1}{8} \int \; d^{4}x \; tr(\epsilon^{\hat{i}\hat{j}\hat{k}\hat{l}}F_{\hat{i}\hat{j}}F_{\hat{k}\hat{l}})\;.
\end{equation}
The momentum in the $5_{th}$ direction is $P_{5}=n/R$. All fields are translation invariant along $x_{5}$ carrying the definite $P_{5}$ momentum and then could not form the representation of $SL(5,Z)$ in the sense of (\ref{sx3s}). Nevertheless, $SL(5,Z)$ symmetry may have the manifestation at the quantum level. Partition function of the $\mathcal{N} = \textnormal{2}$ SYM theory on $T^{4}$, with the contribution from the instanton configurations included, may have the $SL(5,Z)$ invariance, although the explicit calculation is still lacking.

We are especially interested with the 0-mode $A_{I}(t)$ that could be mapped to the membrane configuration. In this case, field strength $ F_{IJ}\equiv H_{IJ5}$ is constant in space. It is reasonable to expect that $H_{IJK}$, which is determined by $H_{IJ5}$, is constant as well. Also suppose the instanton number $n=0$, then (\ref{sx3s}) is well-defined and becomes
\begin{equation}
	H_{0\tilde{i}\tilde{j}}(t)\rightarrow U^{\tilde{l}}_{\tilde{i}}U^{\tilde{m}}_{\tilde{j}} H_{0\tilde{l}\tilde{m}}(t)\;.
\end{equation} 
However, the $SL(5,Z)$ transformation does not always convert the 0-mode $A_{I}(t)$ to another 0-mode $A'_{I}(t)$, which is already manifested in $U(1)$ situation. Consider the on-shell gauge field
\begin{equation}
		A_{0}(t)=0\;,\;\;\;\;\;\;	A_{\hat{i}}(t)=K_{\hat{i}}t\;
\end{equation}
with
\begin{equation}
		H_{0\hat{i}5}(t)=K_{\hat{i}}\;,\;\;\;\;\;\;	H_{0\hat{i}\hat{j}}(t)=\frac{1}{2}\epsilon_{0\hat{i}\hat{j}\hat{k}\hat{l}5}H^{\hat{k}\hat{l}5}=0\;.
\end{equation}
The $SL(5,Z)$ transformed field strength $(H'_{0\hat{i}5}(t),H'_{0\hat{i}\hat{j}}(t))$ may have $H'_{0\hat{i}\hat{j}}(t)\neq 0$, which, cannot be obtained from the space independent $A'_{\hat{i}}(t)$. Intuitively, the $SL(5,Z)$ transformation on $B_{\tilde{i}\tilde{j}}$ acts as 
\begin{equation}
B_{\tilde{i}\tilde{j}}(x_{\tilde{k}},t)\rightarrow U^{\tilde{l}}_{\tilde{i}}U^{\tilde{m}}_{\tilde{j}} B_{\tilde{l}\tilde{m}}(U^{n}_{\tilde{k}}x_{\tilde{n}},t)\;.	
\end{equation}
The components in $B_{\tilde{i}\tilde{j}}$ are not independent but are related via the self-duality relation. For $B_{I5}(t)\equiv A_{I}(t)$ that is constant in space, $B_{IJ}$ is not guaranteed to be constant and so, the $SL(5,Z)$ transformation does not always map the $0$-mode to $0$-mode. Moreover, the transformed $A'_{I}(t)$ may even be $x_{5}$-dependent, making the $SL(5,Z)$ transformation ill-defined. Nevertheless, for some special 0-mode configurations like (\ref{1az}) and (\ref{1q2z}), $SL(5,Z)$ transformation is well defined, making the 0-mode converted into the 0-mode and $M \sim (P_{\hat{i}}, \epsilon_{\hat{i}\hat{j}\hat{k}\hat{l}}W^{\hat{k}\hat{l}}/2)$ transformed into $U M U^{-1}$ for $U \in SL(5,Z)$.

When mapped to the membrane, we may conclude that for the generic membrane configuration, $SL(5,Z)$ symmetry may only be realized at the quantum level with winding modes on $T^{4}$ also taken into account.

\section{Conclusion}

We have studied the U-duality transformation of membrane in $T^{n}\times R$. In lightcone gauge, membrane worldvolume theory is equivalent to the matrix theory, which, with the winding modes on $T^{n}$ taken into account, becomes $(n+1)$-dimensional SYM theory. The original membrane configuration is mapped to the 0-mode of the SYM field. When $n=3$ and $n=4$, $4d$ SYM theory and $5d$ SYM theory are $SL(2,Z)\times SL(3,Z)$ and $SL(5,Z)$ U-duality symmetric. However, the U-duality transformation does not always bring the 0-mode into 0-mode, so the truncation to the 0-mode subspace is not valid. In membrane's respect, this means the duality transformation may convert the configuration without the winding mode into the one with the winding mode. The $SL(2,Z)\times SL(3,Z)$ transformation in $4d$ SYM theory can be realized classically, making the on-shell SYM fields transformed into each other. On the other hand, the $SL(5,Z)$ symmetry in $5d$ SYM theory may only be realized at the quantum level.

\bigskip
\bigskip

\section*{Acknowledgments}

We would like to thank J. X. Lu very much for helpful discussions.
This research was supported in part by the Natural Science Foundation of China under
grant numbers 11135003, 11275246, and 11475238 (T.L.).

\bibliographystyle{plain}

\end{document}